# GENDER GAP THROUGH TIME AND SPACE: A JOURNEY THROUGH WIKIPEDIA BIOGRAPHIES AND THE "WIGI" INDEX

Maximilian Klein (notconfusing.com) max@notconfusing.com
Piotr Konieczny (Hanyang University) piokon@post.pl

## 0 Abstract

In this study we investigate how quantification of Wikipedia biographies can shed light on worldwide longitudinal gender inequality trends. We present an academic index allowing comparative study of gender inequality through space and time, the Wikipedia Gender Index (WIGI), based on metadata available through the Wikidata database. Our research confirms that gender inequality is a phenomenon with a long history, but whose patterns can be analyzed and quantified on a larger scale than previously thought possible. Through the use of Inglehart-Welzel cultural clusters, we show that gender inequality can be analyzed with regards to world's cultures. In the dimension studied (coverage of females and other genders in reference works) we show a steadily improving trend, through one with aspects that deserve careful follow up analysis (such as the surprisingly high ranking of the Confucian and South Asian clusters).

## 1 Introduction

It is an unfortunate but unavoidable fact that encyclopedias always have had a gender bias. One of the dimensions of this bias is the contributors' gender distribution (Thomas 1992; Reagle and Rhue 2011). Just as there were very few women authors among contributors to traditional, printed encyclopedias, recent surveys indicate that women constitute only around 13%-16% of Wikipedia contributors, or Wikipedians (Glott, Schmidt, & Ghosh 2010; Hill and Shaw 2013).

A detailed analysis of Wikipedia editor gender dynamics has been offered by Lam et al. (2011), and the gender disparity of Wikipedia editors has been the subject of mainstream press coverage, itself a subject analyzed in detail by Eckert and Merill (2013). Thanks to a number of studies in the past decade we have arrived closer to understanding why those biases persist in Wikipedia, despite the project's "anyone can edit" nature. This has been attributed to the persisting gender imbalance in computer-related fields, reaching an apex in the Free, Libre and Open Source Software community, where women make up only 1% of participants (Ghosh et al.2002), and with which Wikipedia's community is closely associated (Konieczny 2009).

While the majority of press articles and academic studies have focused on the variations of the research question "Why are so few of Wikipedia's contributors female?", little research has been done to analyze the other aspect of encyclopedic gender gap, namely the skew of *biographical coverage* towards males. This pattern of gender inequality is also prominent on Wikipedia, as indicated by exploratory studies by Lam et al. (2011), Reagle and Rhue (2011), Eom at al. (2014) and Klein (2013a, 2013b, 2014).

In this study we investigate how quantification of Wikipedia biographies can shed light on worldwide longitudinal gender inequality trends. We present an academic index allowing comparative study of gender inequality through space and time, the Wikipedia Gender Index (WIGI).

WIGI is a collection of several new indicators that focus on the ratio of female to male, and nonbinary gender biographies. We analyze its distribution by years of birth and death, "culture" (an aggregated measure of place of birth, ethnicity and citizenship), and language. In the simplest version, we will use the ratio of female biographies indicator, a year of birth indicator, and culture indicator to measure how gender equality is changing.

Though in this study we often discuss the well-known Wikipedia (world's largest and most popular reference work), we rely extensively on data provided by Wikidata. Wikidata is a sister project of Wikipedia, designed to be the machine-readable knowledge base that feeds all Wikipedias. It stores semantic data related to each "item" – the cluster of different language articles that are about the same concept – such as gender or birth dates of individuals. Like Wikipedia, it is collaboratively edited, but it is also curates data that is machine imported from other open datasets around the internet. It is growing exponentially in data and users, and is currently the 4th most popular Wikimedia website by monthly active users, between German and French Wikipedias. Its new popularity is part of the motivation for this research.

## 2 Quantifying the gender gap in Wikipedia's biographies

Wikipedia coverage of women biographies was discussed by Reagle and Rhue (2011). Their exploratory study was focused on the comparative analysis of various reference works rather than of the entirety of the Wikipedia's corpus of biographies. Having asked whether there is a bias in women's representation in Wikipedia biographies, and having compared Wikipedia to Britannica, and both of those works to a number of other English language reference works, the authors observed that Wikipedia, while having a larger total amount of female biographies than Britannica (a result of being many times the size of its competitor), also tends to be less balanced in whom it misses than is Britannica, particularly when it comes to lesser known individuals. More recently, Wagner et al. (2015), focusing on lexical and structural bias in content, suggested that Wikipedia's bias, while certainly present, is smaller than in most other comparable reference works.

Eom et al. (2014) looked at the Top 100 most popular biographical articles at 24 different language Wikipedias. Their study was focused on analyzing popularity of certain topics within Wikipedia, with gender distribution of thereof being only a small digression in their overall study. They observed that female biographies constitute only about 5-10% of the said hundred of biographies. This figure varies with regards to some languages (for example, Finnish number was above average, and Korean, below average), and that the trend suggests a gradual improvement over time, as the ratio of females among the top historical figures improved the closer we got to the modern era. The authors conclude that "most important historical figures across Wikipedia language editions are born in Western countries after the 17th century, and are male."

Finally, the gender gap in Wikipedia biographies have been covered in more detail in informal studies by Klein (2013a, 2013b, 2014). Where preliminary summary analysis showed that Wikipedia female biography ratios were reflective of other encyclopedias' bias, varied by language, and shifting on a year-by-year basis.

## 3 Gender gap indices

We refrain from a thorough methodological analysis of the pros and cons of gender gap indices, sex-disaggregated measures, gender-sensitive aggregate measures and related topics and refer the interested reader to Klasen (2006, 2007), Mills (2010) and Hawken and Munck (2011); however a brief overview of this topic is in order to justify the reason for our proposed approach and to locate our proposed index in the wider frame of reference.

While a number of measures have been proposed in theoretical literature, several major indices have been successfully implemented over a period of years, and are commonly referred to in literature, with no consensus on which is superior (Mills 2010, Hawken and Munck 2011, Beneria and Permanyer 2010). Dreschler, Jutting and Katseli (2008a, 2008b) note that this topic is likely too complex for a single indicator, and recommend a multi-indicator approach for any studies that want to aim for comprehensiveness.

Through one could trace the idea of an index to measure gender-sensitive topics to decades of gender studies literature, the idea has not been successfully implemented until the last few years of the 20th century. Two pioneering measures of gender equality now seen as "traditional" are the UNDP's Gender-related Development Index (GDI) and the Gender Empowerment Measure (GEM), introduced only in 1995. More recently, three new measures were developed: the Gender Equity Index (GEI) introduced by Social Watch in 2005, the Global Gender Gap Index (GGGI) developed by the World Economic Forum in 2006, and the Social Institutions and Gender Index (SIGI) of the OECD Development Centre from 2007.

UNDP's GDI and GEM are both gender-focused extensions of the Human Development Index. GDI's primary focus lies in gender-gaps in life expectancy, education, and incomes.

The GEM, in turn, was designed to measure dimensions omitted from HDI, namely empowerment. It is determined using three basic indicators: proportion of seats held by women in national parliaments, percentage of women in economic decision making positions and female share of income.

While both measures have been and remain widely used, they have been criticized as highly specialized, difficult to interpret, easy to misinterpret, affected by large data gaps, and poorly conceptualized (Dijkstra 2006; Klasen 2006, Dreschler, Jutting and Katseli 2008a, 2008b). Mills (2010) goes as far as to say that "although [GDI and GEM] are often touted as key measures of gender (in)equality, most experts agree that they are in fact not measures of gender inequality at all."

Aiming to provide an alternative to GDI and GEM, the Gender Equity Index (GEI) has been developed to measure all situations that are unfavourable to women. It makes it possible to classify countries and rank them in accordance with a selection of gender inequity indicators in three dimensions, education, economic participation and empowerment. It has been praised for a broader coverage (Mills 2010). However, focusing on socioeconomic opportunities, it has been criticized for ignoring underlying causes of gender inequality such as health (Dreschler, Jutting and Katseli 2008a, 2008b, Mills 2010).

Likely the most widely reported global gender gap index in mainstream press is the World Economic Forum's yearly Global Gender Gap Index. This tool is intended to allow comparative comparison of gender gap across different countries and years. It focuses on four areas of inequality between men and women: economic participation and opportunity, educational attainment, political empowerment and health and survival statistics (Klasen and Schuler 2011). GGGI can be seen as the most comprehensive (Mills 2010), though this has lead to this measure being criticized for being too broad - a criticism previously applied to GDI and GEM as well (Dreschler, Jutting and Katseli 2008a, 2008b).

The most recent attempt to address the perceived inadequacies of those four indices is the Social Institutions and Gender Index (SIGI). It is a composite indicator of gender equality that solely focuses on social institutions (norms, values and attitudes), as well as on the four dimensions of family code, physical integrity, ownership rights and civil liberties. SIGI's authors see it as a tool indented to supplement, not replace, the aforementioned existing measures (Dreschler, Jutting and Katseli 2008a, 2008b). SIGI has been praised for being a valuable measure for developing countries, but criticized as less applicable for the developed ones (Mills 2010).

One of the key limitations of all indices presented here is their reliance on modern statistical data, which reduces their global coverage; even the most broad of those indices rank less than 75% of present-day countries. It also forces them to focus on the last few decades - the period for which such data is available; thus the lack of reliable statistics prevents creation of any index for measuring gender inequality that has a deeper historical perspective (Klasen 2006). Echoing the call of Dreschler, Jutting and Katseli (2008a, 2008b) for the development of indicators able to address other dimensions of the gender inequality, and McDonald (2000) request for a measure with a longitudinal perspective facilitating historical and anthropological studies we therefore propose the following new measure: WIGI (Wikipedia Gender Inequality index). Like SIGI, it is not indented to replace any prior index, but instead we hope it will be a complementary tool, which while limited to only one indicator (ratio of non-males that have a biography for a given time-frame and geographical region) uses Wikipedia's biographies and thus extends the scope of possible analysis throughout the entire recorded human history.

## 4 Research questions:

The preceding literature prompted us to ask the following:

RQ1: Taking year of birth parameter, we can compare the number of Wikipedia's biographies by gender by year (decade, century, millennium). What is the pattern/trend? Can we predict when full equality will be reached?

RQ2: What will be the variations by region/country/nationality/ethnicity/religion/language?

RQ3: What can we learn from variations in variables such as article quality?

## 5 Limitations

Wikipedia is hardly immune to the preexisting sociohistorical bias that affects the entirety of reference and academic works. The results presented here have to be understood in light of the number of constraints.

Almost all of the reference works are affected by gender bias in their contributors' gender and in the biographies listed (except those designed to correct for such biases, an approach which introduces its own set of issues) (Reagle and Rhue's (2011).

Wikipedia's editor base is predominantly male. Thus the fact that our dataset shows that Wikipedia biographies are primarily those of males is a compound result of at least two factors: (a) First, non-males lack of empowerment throughout history, resulting in their lack of access to positions of power or fame that would merit their notability.

(b) Second, an outcome of the Wikipedia's predominantly male editor base unconscious preference to create articles about males and on topics of (at least stereotypical) interest to males, such as military history (Forte, Larco and Bruckman 2009)

Wikipedia, while already the largest reference work created by humankind (already ten times the size of the second largest, i.e. Britannica), is not yet complete, and likely won't be for many years. Data properties are still missing; for example only 89.5% entries have a gender property (see Table 1 for details).

While our data set encompasses biographies from numerous languages, it is affected by the global digital divide (Graham et al. 2014).

Wikipedia has a number of initiatives to reduce the effect of biases listed above, and strives to have a gender-balanced contributor base (Wagner et al. 2015). While encouraging on one level, for our purposes we have to note that such actions inevitably create a counter-bias.

While we operate predominantly in the reference frame of recorded written culture, we have to acknowledge that resulting research will be inevitably biased against cultures which recorded their history orally (Gallert and van der Velden 2014).

No studies, unfortunately, have been carried out to determine how, exactly, the above biases translate into reduced likelihood of a female biography being created, compared to a male or nonbinary biography. While acknowledging that Wikipedia editors are on average more likely to create a male rather than a non-male biography, we have no reason to assume that this is not a bias that holds steadily across time, i.e. there's no reason to assume that the likelihood of a bias introduced by Wikipedia's editors preferences in 20th century biographies is different from the bias in the 7th century biographies. Any resulting bias with regards to the time variable should reflect the content of historical (and contemporary) sources reflecting the progress of women's empowerment (and thus, their presence in such sources) through time.

In order to deal with some of the above problems, we have decided to present data using the nine cultural clusters proposed by Inglehart and Welzel (Inglehart and Welzel 2005): English-speaking, Latin America, Catholic Europe, Protestant Europe, African, Islamic, South Asian, Orthodox and Confucian (see also Image 1). There is, nonetheless, no way to aggregate cultures perfectly. Aggregation in general assumes some loss of fidelity. We acknowledge that those cultural clusters are limited to modern period (year 1500 and later) in history. For instance the notion of having a Protestant and Catholic world before Protestantism and Catholicism is problematic, to say the least; so is the fact that individual born in ancient Greece is classified as Orthodox in this method. Nonetheless due to inevitable correlation between Wikipedia's biographies and world's population, (Pearson = .983**), over 98.5% (N=1,340,754) of biographies with recorded year of birth fall into that period of history. As such, we believe that the use of those cultural clusters does significantly improve the interpretation of the data, aggregating similar patterns from culturally-similar countries, but the reader should keep in mind that the names of Inglehart-Welzel cultural clusters are not historically sound given the time scope of our data (all of human's history).

In lack of any specific data on this topic, this study is based on the assumption that within the scope of limitations discussed above a comparative analysis of women's empowerment through history is nonetheless possible. It is possible that future studies will propose that our data should be modified by weighting it to counteract various the biases, thus modifying our results and conclusions. We look forward to such proposals, and we hope that the open nature of our dataset will make the creation of derivative, weighted indices an easy process.

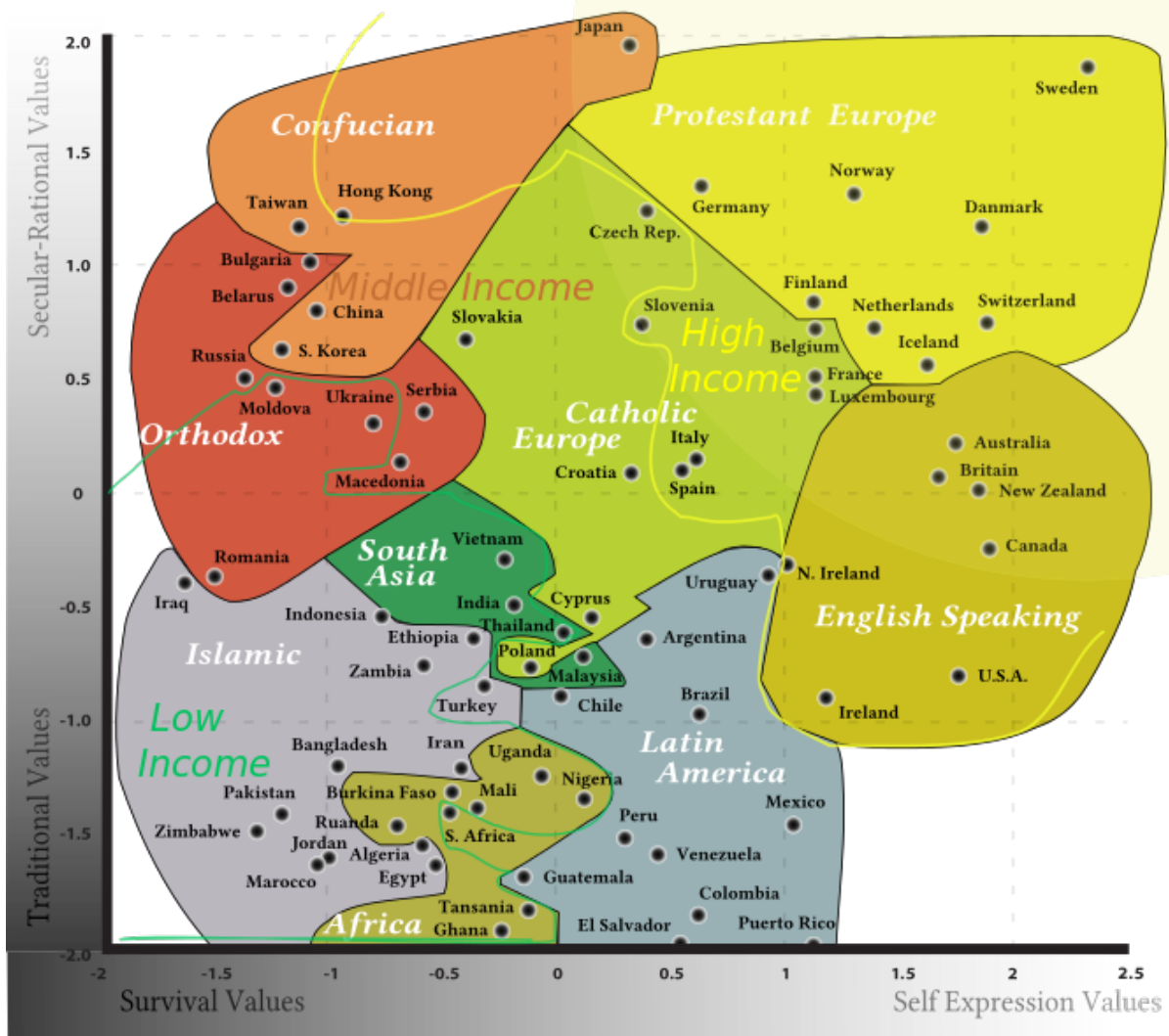

Image 1. Inglehart and Welzel cultural clusters. By DancingPhilosopher [CC BY-SA 3.0 (http://creativecommons.org/licenses/by-sa/3.0)], via Wikimedia Commons

Finally, we would like to stress that the index presented here is intended to allow broad historical comparisons. It is not intended to compete with the other indices which rely on a rich datasets providing statistics on numerous aspects of economy, health, politics, education; instead it is intended as a complementary measure, whose underlying indicators and the open nature of data set they are based on can allow easy incorporation into those indices.

# 6 Methods

This project has been conducted in an Open Notebook Science way, where we have been posting our results and receiving feedback as we work. We worked remotely and had conversations about our in progress research on a wiki page (https://meta.wikimedia.org/wiki/Research:Wikipedia_Gender_Inequality_Index) where other researchers gave us feedback. Nearing the end of our research, we published a blog (Klein 2015) about our preliminary results whereby we received yet more feedback, including corroborating findings from Manske (2015).

## 6.1 Software and Data

In order to work with the most comprehensive database, we have opted to not limit our research to the (most commonly studied) English Wikipedia dataset, but work with data from all 285 different language Wikipedias that existed as of October 2014. We started by obtaining data from the Wikidata database dump. To extract the data from the it we used the Wikidata Toolkit Java Library to subset the entire data into just those items about humans, and then output the results as a CSV (available on our github). The resulting dataset was in turn analyzed using the python-pandas statistical software. Through the quarry tool we queried publicly available Wikipedia database replicas to determine article sizes. We used the pywikibot and *mwparserfromhell* python libraries to get article text for specific biographies during our celebrity-hypothesis testing.

Amazon's *Mechanical Turk* service was utilized to aggregate our list of ethnicities and citizenship. Two independent coders sorted the list into the nine Inglehart-Welzel cultural clusters. For mixed notions, we instructed to use the adjective portion of the ethnic group, e.g. British Raj → English Speaking. Our coders agreed on 68% of cases, and we solved mismatches by hand.

## 6.2 Index construction

Hawken and Munck (2011) work was invaluable in laying out a framework for designing future gender inequality indices.We want to provide the reader upfront with the answers to the following methodological questions:

a) What is the overarching theoretical concept being measured?

All gender gap indices can be understood as "measures of the concept of human development" adjusted for the gender dimension Hawken and Munck (2011). Human development is usually defined as "enlarging people's choices" (United Nations Development Programme 1997:15), and the gender dimension introduces a comparison between different genders.

b) What indicators are selected and how to the connect to the conceptual dimensions of the overarching concept?

We focus on simple and reliable measures of date of birth and culture. One of our indicators - what percentage of different genders have a biography in the target language - can be seen as conceptually similar to the indicator of a having a high-end political and economic position, and correspondingly, the indicator of number of individuals with a Wikipedia biography per unit of time or region is similar to the number of holders of political and economic positions.

All large Wikipedia language editions have a dedicated "Notability" policy for determining which individuals should be included in the scope of the project, through this policy is not identical on all Wikipedias.The English Wikipedia policy states that "A person is presumed to be notable if he or she has received significant coverage in reliable secondary sources that are independent of the subject." What's important for us is the end result: that while some individuals may be notable due to the virtue of their position (member of parliament, royalty, etc.), Wikipedia does not recognize most middle and even senior managers or technicians as notable solely on the virtue of their position. This means that our dataset can be seen as comparable, methodologically-wise, but not identical, to those used in other indices discussed here. A generalization that Wikipedia, like all encyclopedias, writes about "people who are seen as important" should be sufficient, though we (as well as Wikipedia itself) remain conscious of the gender bias present in determining who is important (which is, after all, what we are measuring here).

c) How are the indicator scales designed?

Hawken and Munck (2011) suggest that indicator scales should be consistent with the concept being measured, while offering as much nuance as possible.Our indicator scale is a ratio of gendered biographies to total biographies, categorized by place of birth or citizenship, born within a specific timeframe.

d) How are values assigned to each indicator?

Hawken and Munck (2011) recommend the use of a method assigning values to indicators that is replicable and that generates reliable and valid measures. As our data was collected through data mining, we avoid the most common problems related to subjective measurement such as expert surveys of unknown reliability.

# 7 Results

## 7.1 Descriptive statistics
As of October 14 2014 we inspected a total of 2,561,999 biographies. That is, any Wikidata item with the Wikidata semantic property "instance of: human" ("P31=Q5" in Wikidata vocabulary). On each of those items we look for the following additional properties and found them on the following number of items.

Table 1: Descriptive statistics

|  | % of total | Items with property |
|---:|---:|---:|
| ethnic group | 0.30 | 7,772 |
| country* | 23.47 | 601,361 |
| place of birth | 23.93 | 613,092 |
| date of death | 28.79 | 737,522 |
| citizenship | 41.44 | 1,061,634 |
| culture** | 45.20 | 1,158,086 |
| date of birth | 57.92 | 1,484,003 |
| gender | 89.40 | 2,290,433 |
| at least one site link | 99.05 | 2,537,545 |
| a "Q" ID | 100.00 | 2,561,999 |

**country** is determined by seeing if the *place of birth* is a country, or if it is a city, see if the city has a *country* property.

****culture** is determined by translating *place of birth*, *citizenship,* and *ethnic group* (Wikidata properties P19, P27 and P172, respectively) into one of nine world cultures as per Inglehart-Welzel map of the world. Then we take the consensus of the three aggregated variables. (There were no disagreements between the three variables.)

The first derived statistic of interest is the total gender breakdown. As we've seen above 10.3% is of unknown gender, otherwise we encounter in Wikidata 13.9% female, 75.7% male. We also find 152 cases of nonbinary gender. Normalizing the percentages to only known-gender humans we have 84.4% male, 15.6% female, and $\approx$ 0.0001% nonbinary in Wikidata.

## 7.2 Sanity checking

We carry out a sanity check on whether our data seems to reflect the world at large, by correlating our data with the historical census data and the four other gender indices mentioned. We find that Wikipedia biographies seem to be highly correlated with world's historical population (Pearson correlation coefficient = .983 with $p<0.01$), through unsurprisingly, however, the ratio of Wikipedia's coverage (the percentage of people alive meeting the "notability" criteria) increases as we move closer to the modern times, see Figure 2.

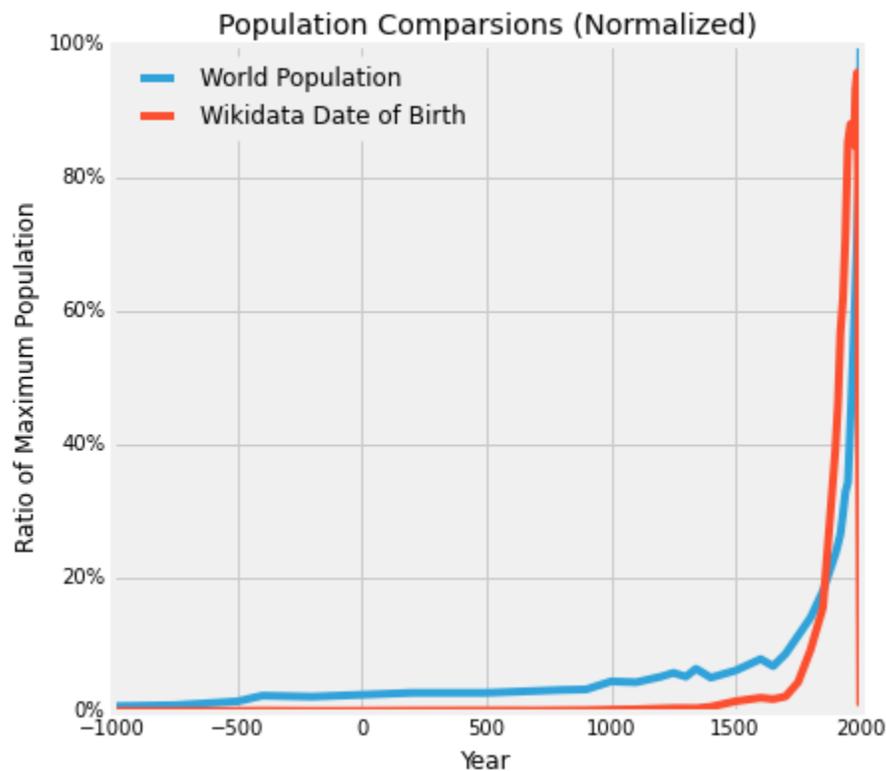

Figure 2: Correlation between Wikipedia biographies and World Population

We next compare WIGI - female ratio by country - to four other indices (GGGI, GDI, SIGI and GEI). Table 2 shows a sample of WIGI and GGGI rankings side-by-side, sorted by WIGI's top 10 rankings. A full comparison between all 5 indexes available on github.

*Table 2. WIGI Top 10[1]*

| Country | GGGI Rank | WIGI Rank | GGGI Score | WIGI Score | Rank Difference |
|---|---|---|---|---|---|
| Sweden | 4 | 1 | 0.8165 | 0.3452 | 3 |
| South Korea | 117 | 2 | 0.6403 | 0.3437 | 115 |
| Philippines | 9 | 3 | 0.7814 | 0.3228 | 6 |
| Bahrain | 124 | 4 | 0.6261 | 0.3171 | 120 |
| Mauritius | 106 | 5 | 0.6541 | 0.2941 | 101 |
| People's Republic of China | 87 | 6 | 0.6830 | 0.2812 | 81 |
| Australia | 24 | 7 | 0.7409 | 0.2760 | 17 |
| Japan | 104 | 8 | 0.6584 | 0.2732 | 96 |
| Nicaragua | 6 | 9 | 0.7894 | 0.2727 | -3 |
| Swaziland | 92 | 10 | 0.6772 | 0.2593 | 82 |

---

1 Full dataset is available at https://github.com/notconfusing/WIGI/blob/master/helpers/foreign_indexes/WIGI_comparison.csv

Next, for each of the four alternate indices we use the national rankings to produce the Spearman Rank correlation statistic between the two rankings. Then we perform a calibration step to find the starting decade to use in subsetting WIGI which produces the highest correlation with the alternative index. Table 3 shows four calibration results.

| Table 3. National-WIGI compared to Alternative indices | | | |
|---|---|---|---|
| **Index** | **Spearman Correlation** | **Significance** | **Calibrated Start Decade** |
| GEI | 0.417 | p<0.001 | 1910 |
| SIGI | 0.338 | p<0.001 | 1910 |
| GGGI | 0.310 | p=0.03 | 1890 |
| GDI | 0.278 | p<0.001 | 1910 |

Each alternative index shows some statistically significant moderate correlation the our WIGI measure. This proves that the female composition of Wikidata items of humans associated with a country can be a helpful tool in enhancing gender inequality indices.

Additionally the fact that each alternative index most highly correlates when we consider only those biographies starting around 1900 is a positive sanity check for our data in the light of the fact that traditional indices talk about modern history.

Finally we also present the results of the WIGI measure as a geographic map. See figure 3.

**WIGI Visualization**

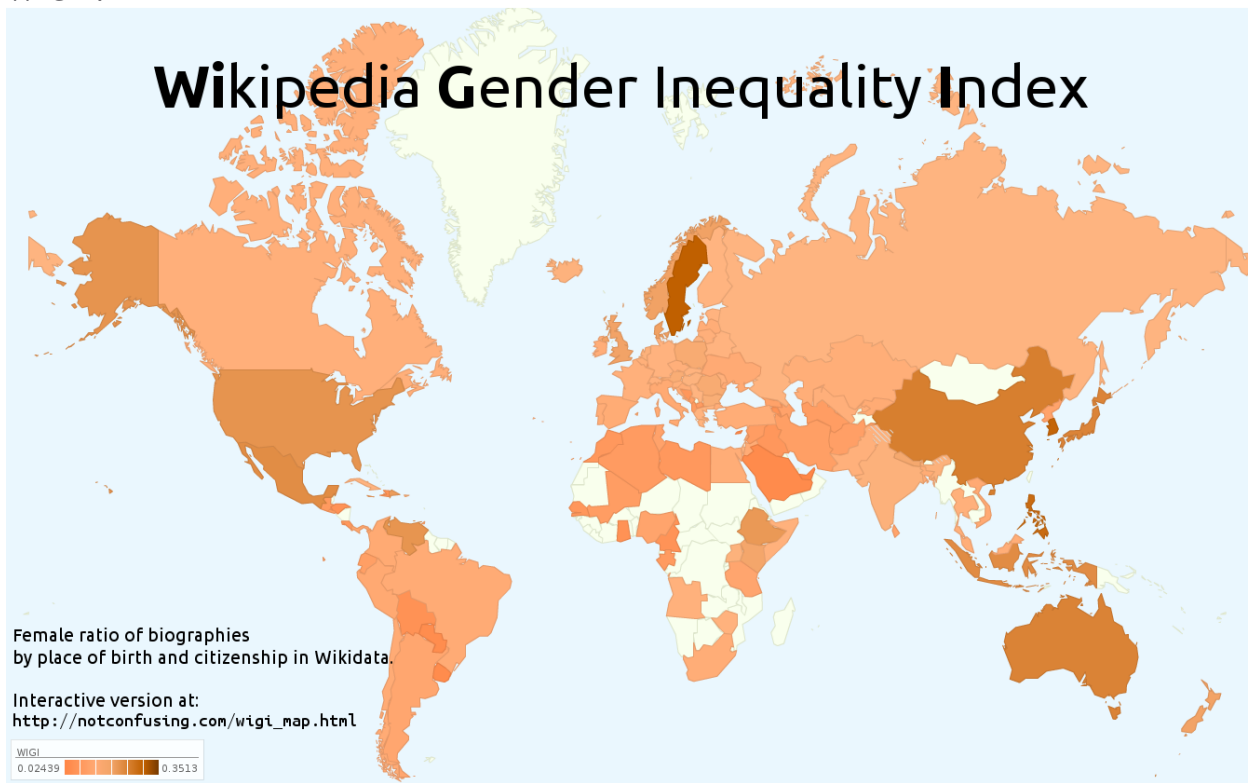

Figure 3: WIGI index world map

## 7.4 Gender Ratios Over Time

We start looking at how the *ratios* of female and nonbinary genders develop over time in Figure 4. We adjust our viewing window here to start at 1400CE here because the data is otherwise too sparse to provide meaningful visual data. We aggregate the nonbinary genders into a single class for the ease of visualization..

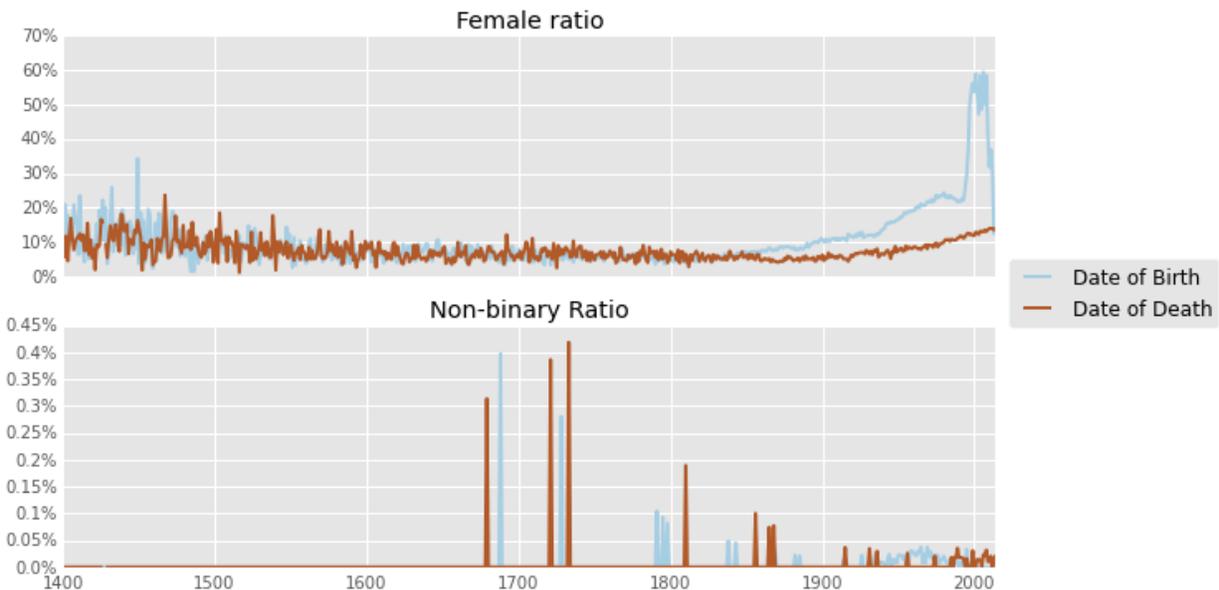

Figure 4. Composition of Wikidata genders in modern history.

Since about 1800 to present, the female ratio of biographies is greater when using the date of birth measure than the date of death measure. In other words, we find that recording female date of birth has become more prominent than date of death.

Even with discounting very recent trends of the last 20 years, which describe humans that are just entering adulthood or younger, the female ratio by date of birth is rising exponentially. Although it may not necessarily indicate equity, fitting an exponential model to this rise in ratio we can calculate when the female percentage would reach approximately 50%. That model giver the equation $femaleratio(year) = 0.9871e^{-1.1059(year)+2309.1685} + 0.0383$, and solving it yields that it would be February 2034 when the exponential extrapolation would reach 50% female representation. We suspect that in reality we will encounter a logistics model - not an exponential model - but presently we haven't encountered the inflection point of slowing rate of growth yet.

## 7.5 Gender Ratios By Culture

We make a cross-tabulation of gender by our aggregated culture measure. A Chi-squared test shows the observed distributions of gender by culture to vary significantly (p<0.01). We graph the female percentage of biographies by culture in Figure 5.

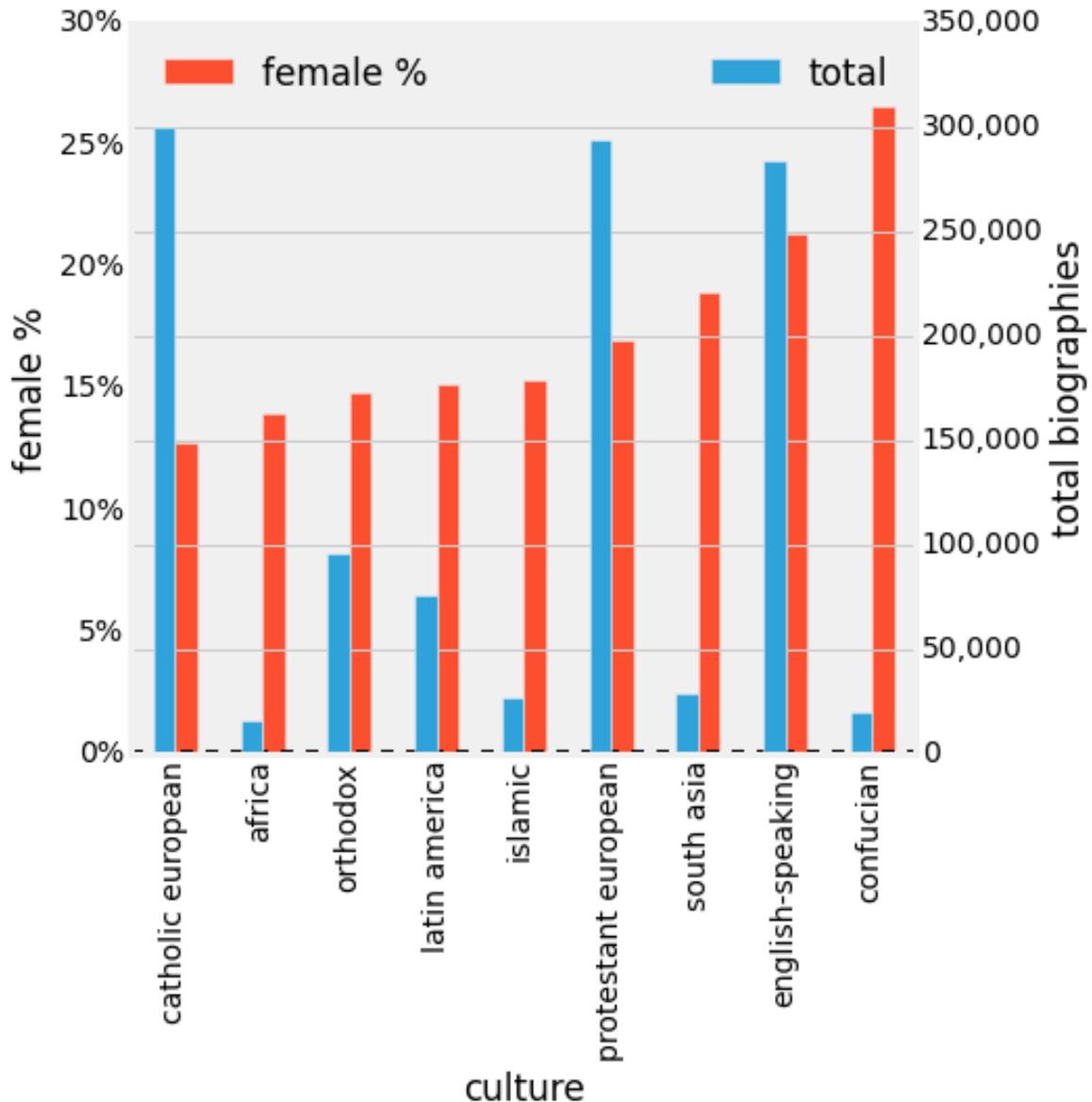

Figure 5. Percentage of female biographies by culture.

We find a large difference in absolute number of biographies by culture. It might be that European and English-Speaking biographies are simply more likely to be described in Wikidata at the moment, as an artefact of the volunteer import process.

Still, although European and English-speaking world dominate in total items, they perform differently in female ratio. Inspecting the female ratio as-is, we find a very high showing for the Confucian culture. To explain it we propose a hypothesis that this is because the phenomenon of celebrity is larger in those cultures, and celebrity is more evenly gender-distributed.

Next, we provide the same graph for nonbinary percentages of biographies by culture in Figure 6. The cultures are ordered in the same way as the female graph for ease of comparison. Notice that the ordering of total items is relatively similar to the female graph – which suggests that similar factors affect recording of female biographies and those of nonbinary gender individuals.

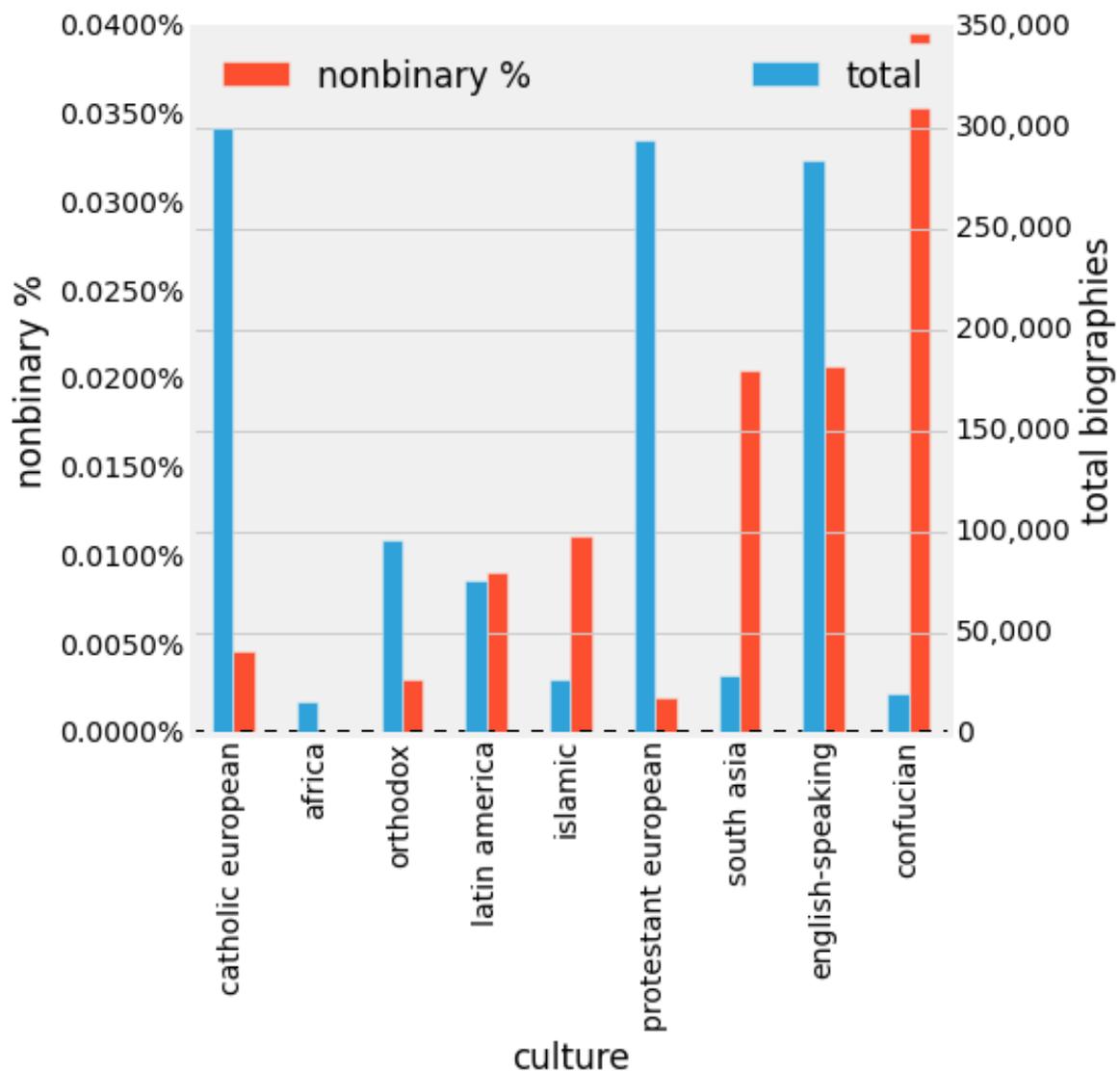

Figure 6 Percentage of nonbinary biographies by culture.

The original Wikidata policy stated that gender assignment "must be one of 'male', 'female', or 'intersex'", but has since changed with community discussion. Currently Wikidata is allowing an increasing list of values and we found occurrences of transgender female (89), transgender male (20), intersex (13), genderqueer (7), fa'afafine (1), and kathoey (1). Therefore in allowing Wikidata to be descriptive rather than being prescriptive we have seen our dataset become richer. We note that we see a 4.45:1 proportion of transgender females to transgender males, which is similar to the 3:1 reported in (Landén 1996). And similarly that 89 transgender female of about 2,500,000 items is about 1:30,000 ratio, which a number that is typically reported in mainstream media (Conway 2001).

## 7.6 Gender Ratios Over Time

Next, we combine the three sets of variables - gender, culture and time. To note our sample size as we continue, only 951,101 or about 35% of total records have all of date of birth, culture, and gender data.

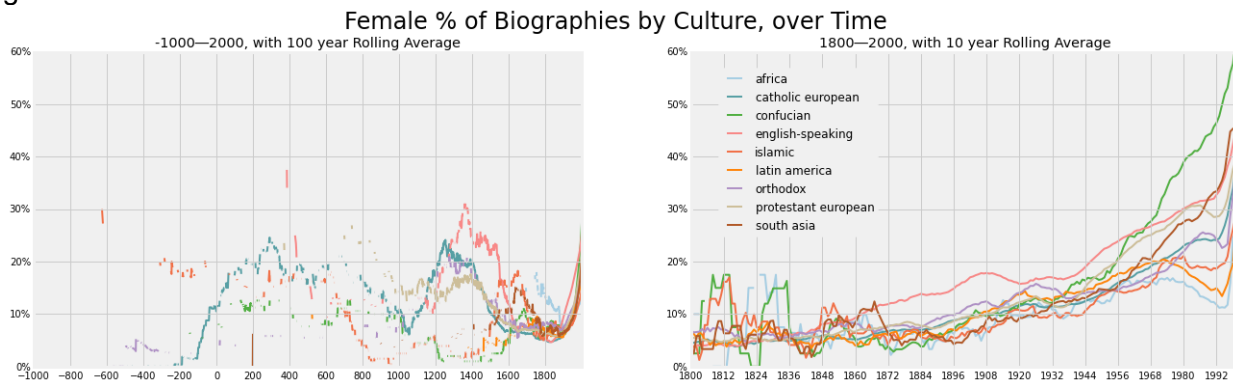

Figure 7. Gender ratios over time and culture.

In Figure 7 we see that the recent past around 1800 is a low point for female recognition in all cultures and most of recorded history. Likewise visually it is evident that historical trends in different cultures have peaked at much higher ratios than one or two centuries ago. In the modern historical graph, we can see a rise occurring for all cultures, and super-linear growth even for the Confucian and South Asian countries .The sky-rocketing ratios after 1990 are less significant as noted above, due to inherent non-notability of young individuals.

## 7.7 Gender by Wikipedia Language

Now let us recall that there is one more dimension we have recorded, the sitelink dimension, which indicates whether or not for an item a Wikipedia language has an entry for it. To be clear, say for instance that Finnish Wikipedia has an article about a Japanese human; the sitelink dimension records this as comment on Finnish Wikipedia. With this data we can analyse the female and nonbinary tendencies of a Wikipedia language, rather than a nationality or culture. Figure 8 shows the relative frequencies of female articles per Wikipedia language, versus the size of the language.

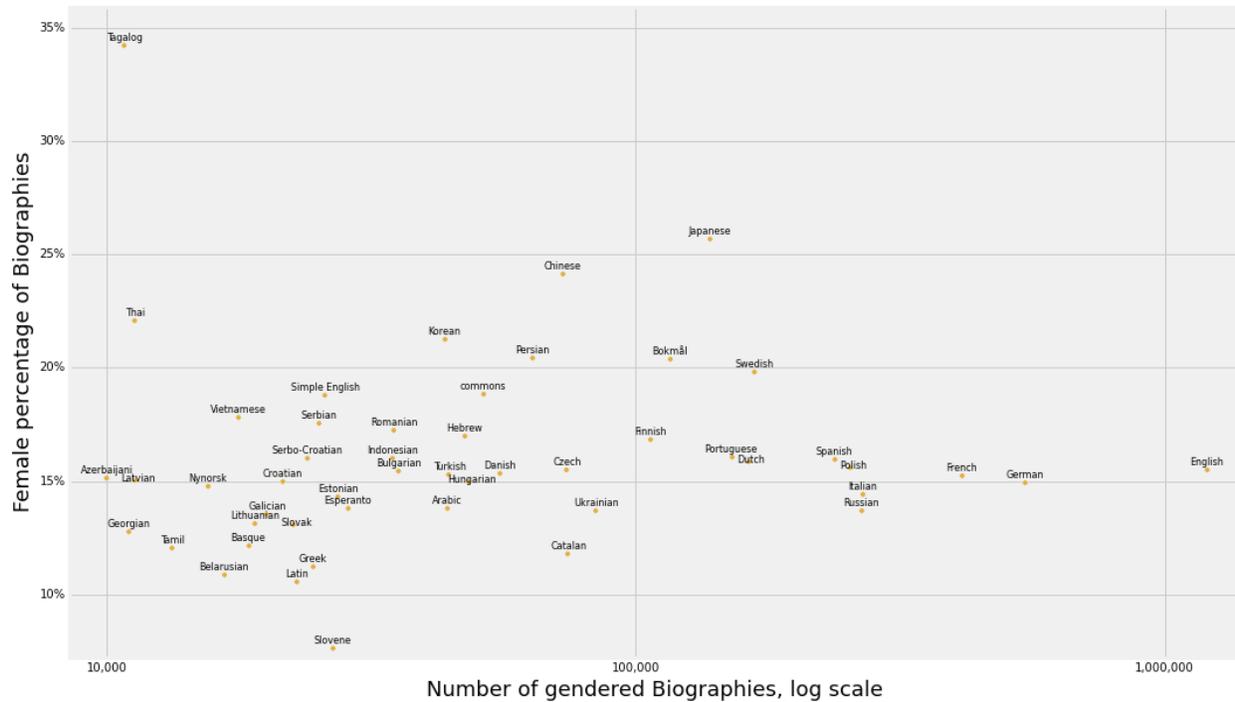

Figure 8. Percentage of female biographies by language of Top 50 Wikipedias.

The visual technique we use is to look at for the points whose magnitude from the origin is greatest. In general there is no simple trend linking Wikipedia size to female representation. We see relatively a flat constant rate, with a few Wikipedias standing out, like the Japanese, Chinese and Tagalog. So again we are seeing some evidence for the languages of Confucian and South Asian cultures being less gender biased.

We repeat the analysis for nonbinary humans in Figure 9.

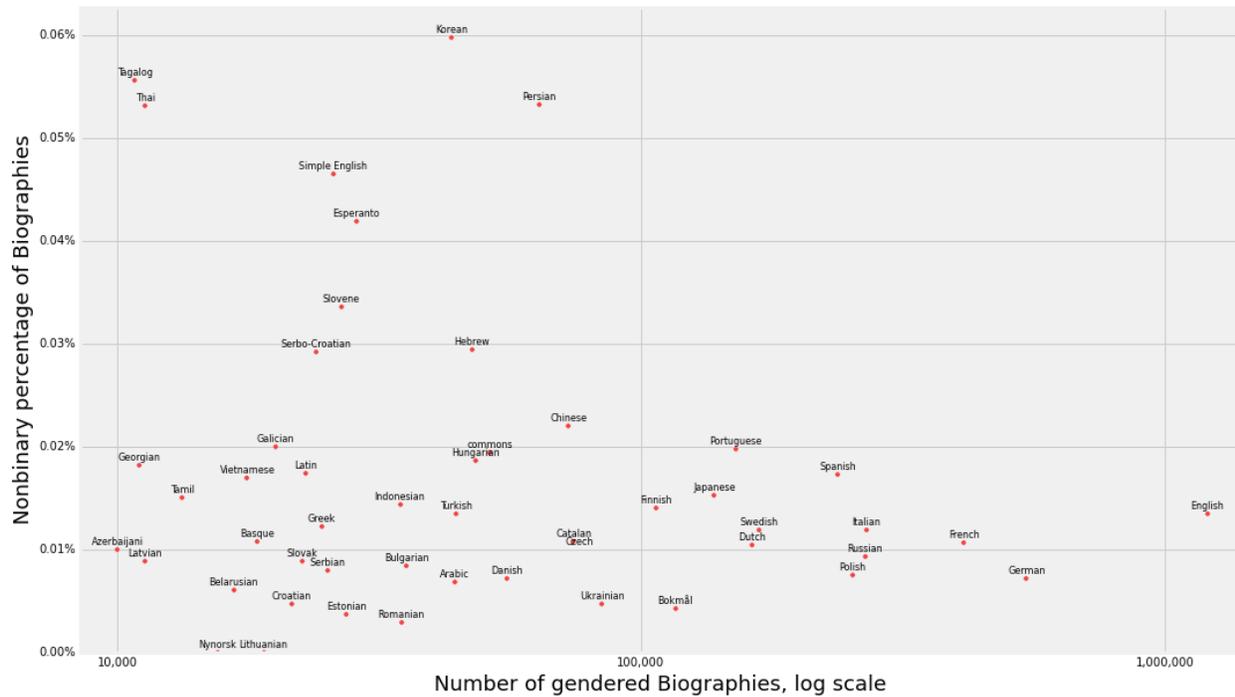

Figure 9. Percentage of nonbnary biographies by language of Top 50 Wikipedias.

With only 152 data points the reliability of this analysis is low, but we again see the languages of Confucian and South Asian countries towards the top.

## 7.8 Language-uniqueness

Viewing by-Wikipedia gives us an overview of how a Wikipedia treats gender, but only part of each language's culture shines through as each language has articles about world biographies as well as local biographies. We now turn to separating articles into those articles which exists in only one language - which we call "language-unique" - and those that exist in more than one language - which we call "language-many". We then compute a measure, graphed in Figure 10, which is the difference between language-unique and language-many female ratios, which describes how how much more or less female-oriented a language is when talking about its "local heroes".

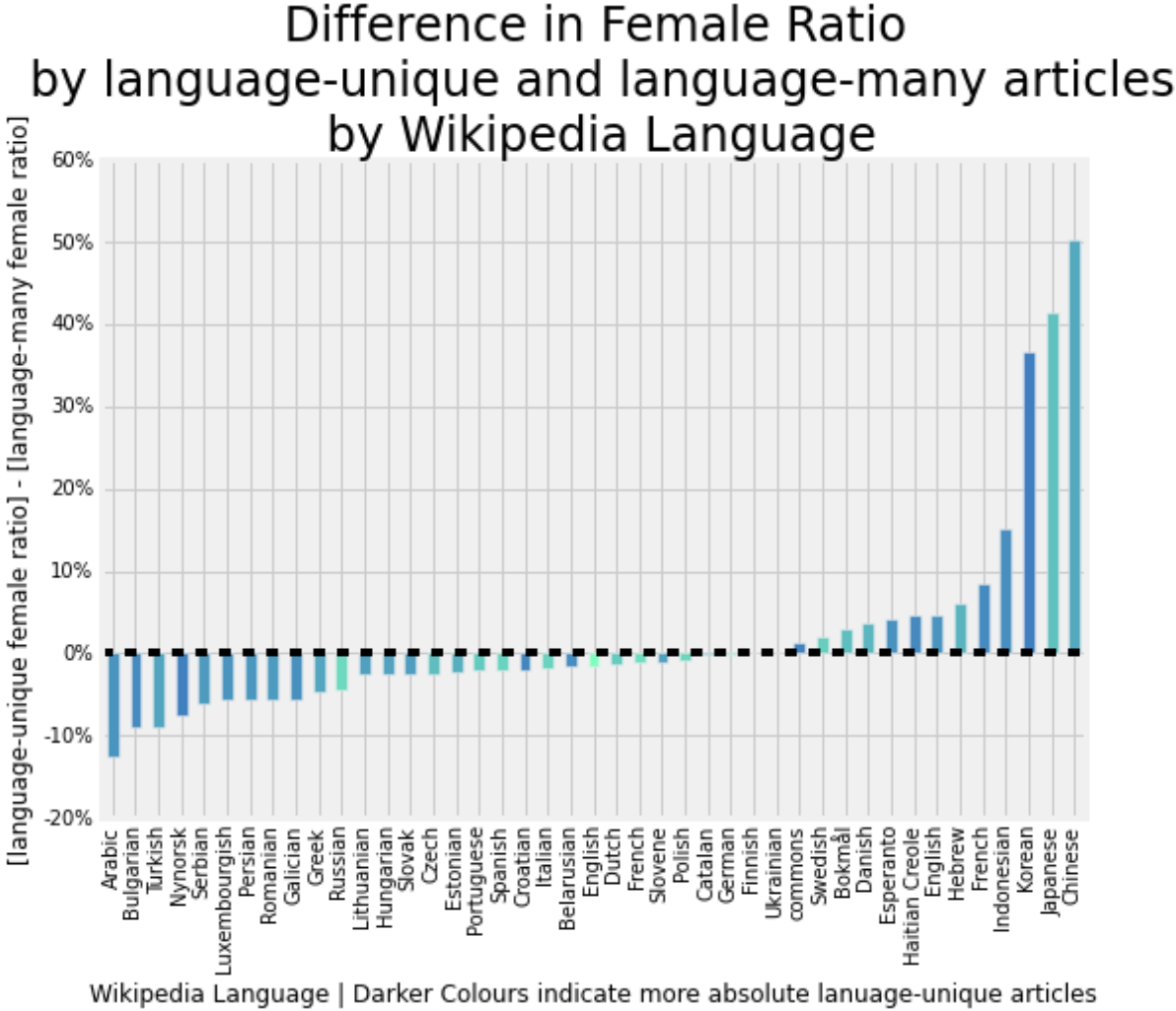

Figure 10. Difference in female ratio by language-unique and language-many articles by language of Wikipedia.

Notice we also provide a colour indicator which displays the absolute number of language-unique articles (darker is larger). So again in broad strokes which see that Confucian South Asian and also Nordic nations seem to focus to write more female-oriented local hero articles.

## 7.9 Gender by Aggregated Wikipedia Language

To sure up the idea of cultural influence in the sitelinks analysis we aggregate the languages into the nine World Cultures as before. In this case, since there are only 285 languages, we assigned all of the languages by hand, rather than resorting to Mechanical Turk. Additionally we created a separate category for the constructed-language Wikipedias.

To clarify, the technique used here is that every Wikidata item counts towards a culture if a sitelink exists in at least one language associated with that culture. So if an article has language links to English, Chinese, and Japanese wikipedia, that item counts only once towards each of the English-speaking and Confucian categories

.

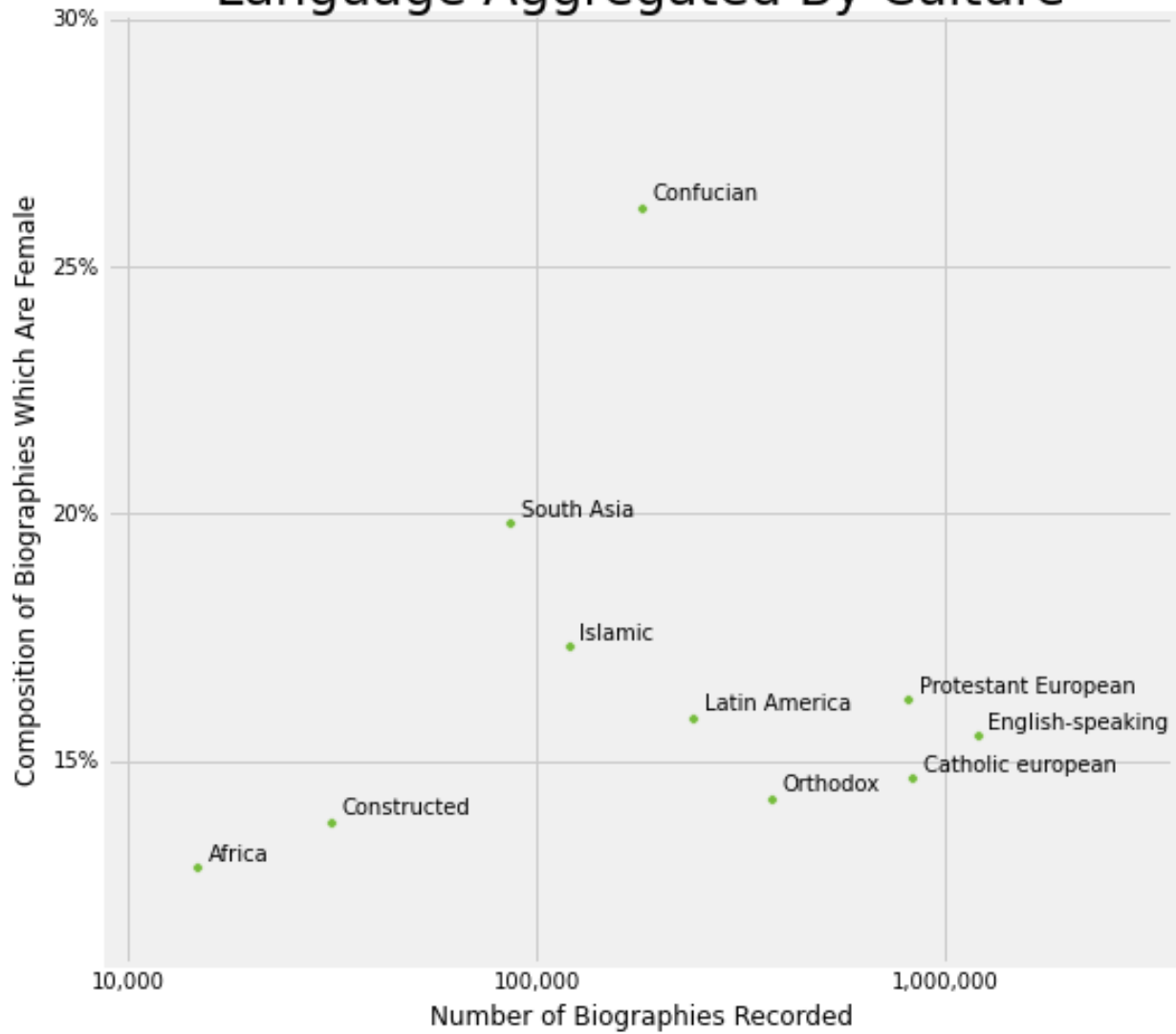

Figure 11. Percentage of female to total biographies by culture.

We now see the Confucian and South Asian cultures confirmed as the top two cultures in terms of female biography ratio. We also see the European / English-speaking cultures clustering very closely to each other, showing not much difference either in total biographies or in female ratio. Islamic, Latin American and Orthodox cultures show an inverse relation between size and female ratio, but that is not consistent with the the African and Constructed-Languages trend.

## 7.10 Mean Article Size

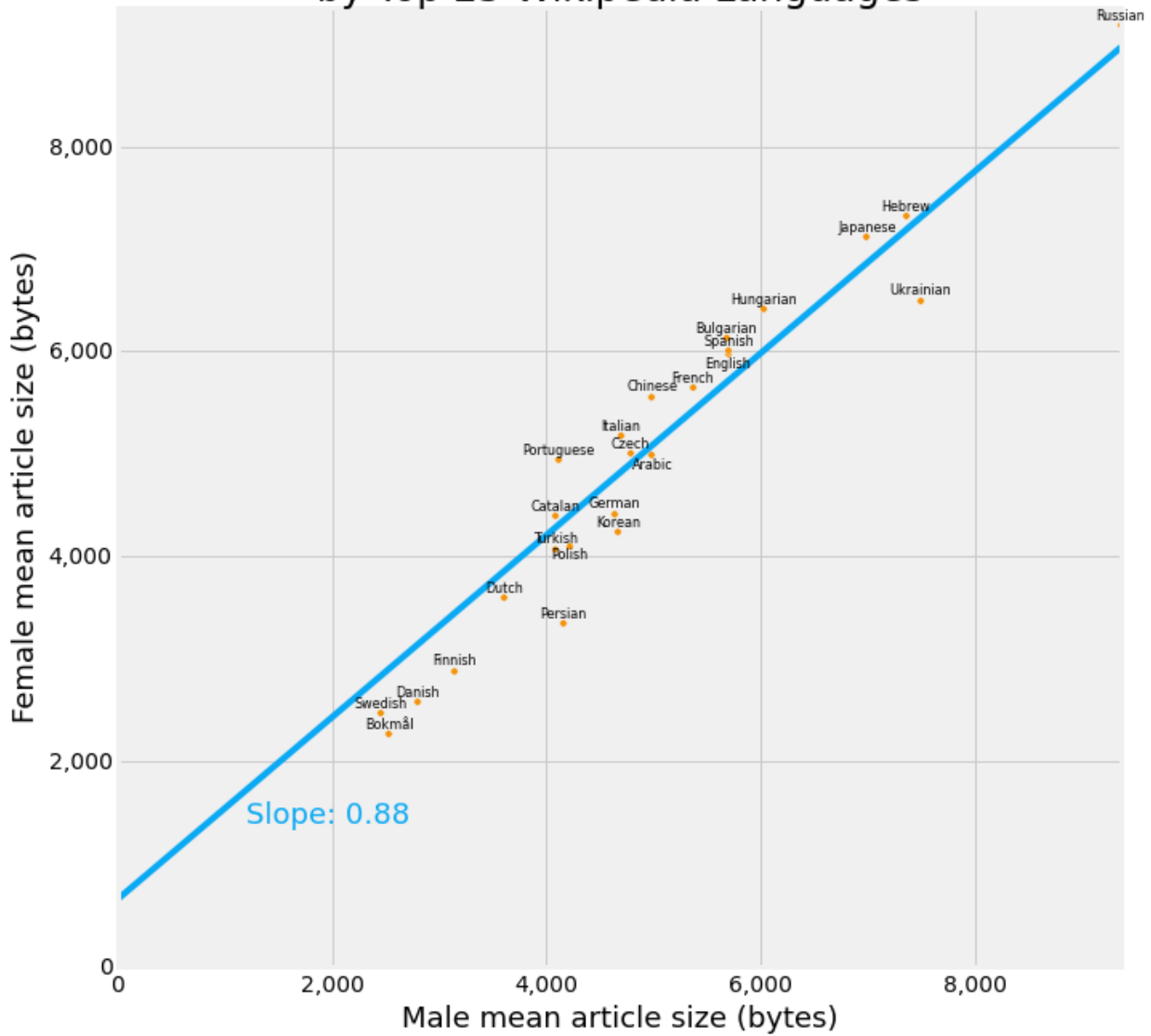

Figure 11. Trend of ratio of mean article size by gender for Top 25 Wikipedias by language.

We now turn our attention to another measure of a Wikipedia's size - the mean number of bytes per biography article (a method proposed by Wikimedian Magnus Manske). We then divide our data by binary gender. In Figure 11 we plot the male mean article size on the x-axis, and the female mean article size on the y-axis for the top 25 Wikipedias by number of biographies. Further we fit a linear regression to the total data across all languages which produces an $R^2$ fit of 0.844, and then plot it as shown. The slope of the linear model is 0.88, (with intercept 648), which indicates that in general the mean article size quite consistently across all languages is about 10% less for female articles. Note also that languages which use unicode characters will use two-bytes per character, so mean byte size may not be a direct information-size comparison. The goodness of fit for this analysis is high, and amongst the top 25 languages no language specifically stands out by deviating largely from the linear model.

## 7.11 Celebrity Hypothesis

Now we have a more coherent picture about which types of Wikipedias by language are focusing on female articles.

We investigate our celebrity hypothesis, that languages with higher female ratios are also more celebrity-focused. For the Chinese, Japanese, Korean, Tagalog, Urdu (which displayed high female ratios) and German and English Wikipedias (to act as baselines), we retrieved the page content of each Biography from 1930 until 1989. We used 1989 as a cut off to avoid any skewing from the reasons why someone born after 1990 might be notable.

We declare for the English or foreign language words that are associated with *celebrity*. The dictionaries used are described in Table 4:

Table 4. Celebrity terms tested.

| Language | Celebrity Terms |
| --- | --- |
| Japanese Wikipedia | '俳優', '選手', '歌手', 'ミュージシャン', 'モデル', 'アイドル' |
| Chinese Wikipedia | '演員', '運動員', '歌手', '音乐家', '模特兒', '偶像' |
| Korean Wikipedia | '□□', '□□', '가□', '□□가', '□□', '□□' |
| Tagalog Wikipedia | 'artista', 'aktor', 'player', 'mang-aawit', 'musikero', 'modelo', 'idolo' |
| Urdu Wikipedia | 'اردو', 'کھلاڑ', 'گلوکار' , 'موسیقار' , 'ماڈل', 'بت' |
| German Wikipedia | 'schauspieler' , 'spieler', 'Musiker', 'Sänger', 'Modell', 'Idol' |
| English Wikipedia | 'actor', 'actress', 'player', 'singer', 'musician', 'model', 'idol' |

We prefered to search the English language article if it was available, and otherwise searched the foreign language content. A celebrity was defined as a biography that contains one of the above words within the first 200 characters of its Wikipedia entry.

We show a heatmap comparing the language, the decade and, the gender, and celebrity percentage in Figure 12.

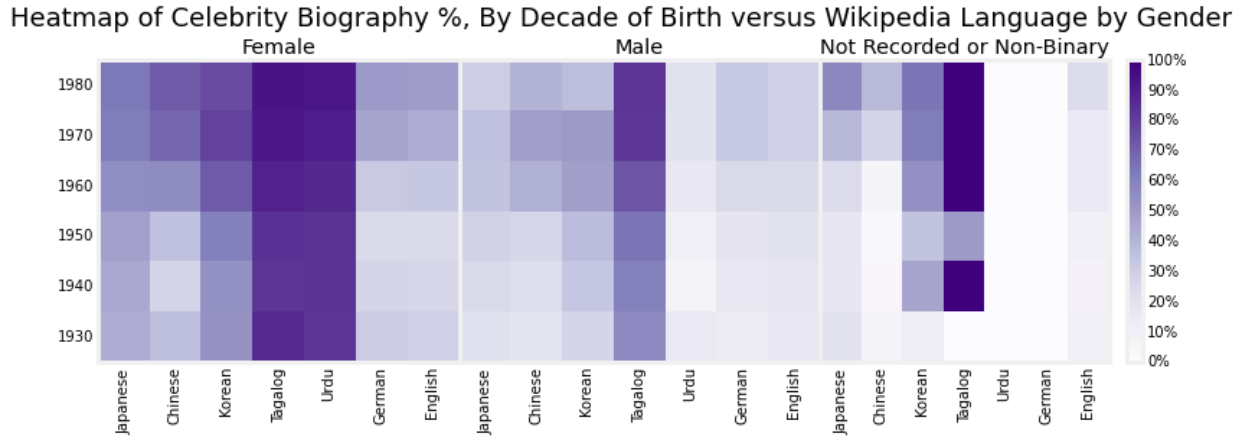

Figure 12. Heatmap of percentage of celebrity biographies by decade of birth versus Wikipedia language by gender.

We can see that the female matrix is darker in general that the other two matrices, as recorded females are more likely to be celebrities among these languages. Likewise one can see that in general the heatmap transitions to being darker at the top than bottom, which describes increase in the percentage of celebrities in most languages in recent years.

Lastly we observe some vertical-striped features showing that some languages, in particular Tagalog, include more celebrities across gender and time.

To determine the significance of the effects we perform a logistic regression analysis in predicting the celebrity percentage variable. The coefficient matrix is shown in Table 5.

Table 5. Celebrity logistic regression

| Celebrity Gender Logistic Regression Coefficient Matrix | | | |
|---|---|---|---|
|  | coef | z | P>\|z\| |
| enwiki | 0.0509 | 0.058 | 0.954 |
| jawiki | 0.7763 | 0.927 | 0.354 |
| kowiki | 1.3834 | 1.662 | 0.097 |
| tlwiki | 3.0009 | 3.176 | 0.001 |
| urwiki | 0.8901 | 1.025 | 0.306 |
| zhwiki | 0.5383 | 0.637 | 0.524 |
| female | 1.3580 | 2.999 | 0.003 |
| decade | 0.0236 | 1.823 | 0.068 |
| intercept | -47.9056 | -1.888 | 0.059 |

The *female,* and *Tagalog*, variables are significant predictors of celebrity with p<0.05. With only a slightly higher significance threshold, *decade,* and *Korean* also become predictors. This lends a credence to the notion that in the cases in which women are recorded in Wikipedias, they have a strong tendency to be a celebrity, and that in general over time Wikipedias are becoming more celebrity focused.

# 8 Discussion

**RQ1: Taking year of birth parameter, we can compare the number of Wikipedia's biographies by gender by year (decade, century, millennium). What is the pattern/trend? Can we predict when full equality will be reached?**

In response to RQ1 we compared the number of Wikipedia biographies by gender and by date of birth and death. We found a difference in the gender ratios between date of birth and date of death. These ratios have been higher in ancient history than in some modern history, but rely on sparse data. In the modern age, especially since a globally identified low point of the 19th century, female ratios have been rising by both date of birth and death measures. Although both female ratios are rising, somehow date of birth is outstrips date of death. A possible interpretation of this fact is that we are viewing an artefact of a social tendency that females are notable by birth because of their familial status (e.g. royal family, or daughter of famous person), but recording date of death has more to do with earning lifetime notability in world a biased world. Still, female ratios are rising, and at an exponential rate. Using the current exponential rate we calculate that in 2034 binary gender ratios will reach numerical parity.

**RQ2: what will be the variations by region/country/nationality/ethnicity/religion/language?**

With our data split by our "culture" dimension, we see cultures range independently of their size from about 12% to 27% female biography representation over time. We also saw a similar relation between representation of nonbinary genders and female representation.

When viewing our culture dimension longitudinally we see very different ancient-historical trends among different cultures. For instance we see the Catholic-Europe having an early peak around 200AD, later around 1200, which is followed by the English-speaking culture peaking around 1300 to around 30%.

Approaching modern history, around 19th century all cultures seemed to have a low flat female ratio, but have been rising since. In that rise we find the female ratio increasing in Confucian and South Asian cultures more quickly than others, followed English-speaking and European cultures where the ratio is growing but at slower rate. Islamic, Latin American, and Orthodox cultures have also been improving their ratio but data suggests that in most recent years growth has been stagnant or slightly declining.

RQ3: What can we learn from variations in variables such as article quality?

We looked at correlations between female ratio and: the size of a Wikipedia by number of gendered articles, the mean size of all gendered articles in bytes, and celebrity ratio.

We found that there is no simple trend linking number of gendered articles of a Wikipedia to female ratio. Likewise even when we aggregated our languages into world cultures there was not an obvious way that total number of articles relate to female ratio. Yet there was a significant constant relationship between a the mean article sizes by binary genders across all languages.

Taken together, these three findings show the how gender-bias works more at a macro level. Even while each culture or language's female ratio remains unrelated to it's size, the article-length bias remains. Perhaps a culture may include more female biographies articles, but it will not mean that the attention paid to each biography is raised as well.

In exploring the celebrity hypothesis we found that the language of a Wikipedia can be a significant predictor in determining article celebrity tendencies. Moreover females in our test data were significantly more likely to be celebrities, and celebrity is rising with date of birth year.

From the perspective of nonbinary genders, we find evidence that our data lines up with baseline mainstream statistics about nonbinary genders. As our transgender ratio at 4.45:1 and, population prevalence at 1:30,000, we find that Wikipedias are reflecting a popular, albeit perhaps biased, view of the world. While this lends support to showing our index measures as being accurate, it also confirms that the Wikipedia view of the world is not unbiased.

# 9 Conclusion

Is this bias of Wikipedia simply a reflection of our gender-unequal reality, or is it also a contributor issue? Because Wikipedia is embedded in so many of our daily activities and routines, it is not just a reflection of the world – but it also a tool used to produce it (Graham et al. 2012). As Wikipedia shows that percentage of notable women increases through time, another metric should be possible, based on Reagle and Rhue (2011) work: one measuring when Wikipedia will stop being biased against inclusion of women, in other words – when there is no statistically significant difference between whether a missing biography is male or female. Such a metric may also help design a set of weights for a refined version of this index as proposed above.

Our research confirms that gender inequality is a phenomenon with a long history, but whose patterns can be analyzed and quantified on a larger scale than previously thought possible. Through the use of Inglehart-Welzel cultural clusters, we show that gender inequality can be analyzed with regards to world's cultures. In the dimension studied (coverage of females and other genders in reference works) we show a steadily improving trend, through one with aspects that deserve careful follow up analysis (such as the surprisingly high ranking of the Confucian and South Asian clusters).

We hope that this study proves that Wikipedia and Wikidata, through certainly affected by a number of limitations, can be used to uncover previously unknown social patterns. The proposed WIGI index should be a useful supplemental measure to enhance other gender inequality indices, as well as providing a quantitative measure to analyze this topic for time periods and regions not covered by more traditional datasets.